\input harvmac

\noblackbox


\lref\strjunc{
O.~Aharony, J.~Sonnenschein and S.~Yankielowicz, ``Interactions of
strings and D-branes from M theory,'' Nucl.\ Phys.\ B {\bf 474},
309 (1996) [arXiv:hep-th/9603009];
J.H. Schwarz, ``Lectures on Superstring and M theory Dualities:
Given at ICTP Spring School and at TASI Summer School", {\it Nucl.
Phys. Proc. Suppl.} {\bf 55B} (1997)1, hep-th/9607201;
 O.~Aharony and A.~Hanany,
``Branes, superpotentials and superconformal fixed points,''
Nucl.\ Phys.\ B {\bf 504}, 239 (1997) [arXiv:hep-th/9704170];
M.~R.~Gaberdiel and B.~Zwiebach, ``Exceptional groups from open
strings,'' Nucl.\ Phys.\ B {\bf 518}, 151 (1998)
[arXiv:hep-th/9709013];
O.~DeWolfe and B.~Zwiebach, ``String junctions for arbitrary Lie
algebra representations,'' Nucl.\ Phys.\ B {\bf 541}, 509 (1999)
[arXiv:hep-th/9804210];
O.~Aharony, A.~Hanany and B.~Kol, ``Webs of (p,q) 5-branes, five
dimensional field theories and grid  diagrams,'' JHEP {\bf 9801},
002 (1998) [arXiv:hep-th/9710116];
K.~Dasgupta and S.~Mukhi, ``BPS nature of 3-string junctions,''
Phys.\ Lett.\ B {\bf 423}, 261 (1998) [arXiv:hep-th/9711094];
A.~Sen, ``String network,'' JHEP {\bf 9803}, 005 (1998)
[arXiv:hep-th/9711130];
S.~J.~Rey and J.~T.~Yee, ``BPS dynamics of triple (p,q) string
junction,'' Nucl.\ Phys.\ B {\bf 526}, 229 (1998)
[arXiv:hep-th/9711202];
O.~Bergman and B.~Kol, ``String webs and 1/4 BPS monopoles,''
Nucl.\ Phys.\ B {\bf 536}, 149 (1998) [arXiv:hep-th/9804160].
}

\lref\BP{
R.~Bousso and J.~Polchinski, ``Quantization of four-form fluxes
and dynamical neutralization of the  cosmological constant,'' JHEP
{\bf 0006}, 006 (2000) [arXiv:hep-th/0004134].
}

\lref\HorowitzNW{ G.~T.~Horowitz and J.~Polchinski, ``A
correspondence principle for black holes and strings,'' Phys.\
Rev.\ D {\bf 55}, 6189 (1997) [arXiv:hep-th/9612146].
}

\lref\domainwalls{
S.~Gukov, C.~Vafa and E.~Witten, ``CFT's from Calabi-Yau
four-folds,'' Nucl.\ Phys.\ B {\bf 584}, 69 (2000) [Erratum-ibid.\
B {\bf 608}, 477 (2001)] [arXiv:hep-th/9906070];
S.~Kachru, X.~Liu, M.~B.~Schulz and S.~P.~Trivedi, ``Supersymmetry
changing bubbles in string theory,'' JHEP {\bf 0305}, 014 (2003)
[arXiv:hep-th/0205108].
}

\lref\tometal{
T.~Banks, ``A critique of pure string theory: Heterodox opinions
of diverse  dimensions,'' arXiv:hep-th/0306074;
T.~Banks and W.~Fischler, ``An holographic cosmology,''
arXiv:hep-th/0111142.
}

\lref\bfss{
T.~Banks, W.~Fischler, S.~H.~Shenker and L.~Susskind, ``M theory
as a matrix model: A conjecture,'' Phys.\ Rev.\ D {\bf 55}, 5112
(1997) [arXiv:hep-th/9610043].
}

\lref\BL{
D.~Berenstein and R.~G.~Leigh, ``String junctions and bound states
of intersecting branes,'' Phys.\ Rev.\ D {\bf 60}, 026005 (1999)
[arXiv:hep-th/9812142].
}

\lref\SW{
L.~Susskind and E.~Witten, ``The holographic bound in anti-de
Sitter space,'' arXiv:hep-th/9805114.
}

\lref\AcharyaII{ B.~S.~Acharya, F.~Denef, C.~Hofman and
N.~Lambert, ``Freund-Rubin revisited,'' arXiv:hep-th/0308046.
}

\lref\moore{G. Moore, in progress}

\lref\alexetal{W.-Y. Chuang, A. Saltman, E.Silverstein, in
progress}

\lref\Douglas{
S.~Ashok and M.~R.~Douglas, ``Counting flux vacua,''
arXiv:hep-th/0307049.
}

\lref\SKunp{S. Kachru, unpublished}

\lref\KKLT{
S.~Kachru, R.~Kallosh, A.~Linde and S.~P.~Trivedi, ``De Sitter
vacua in string theory,'' arXiv:hep-th/0301240.
}

\lref\MSS{
A.~Maloney, E.~Silverstein and A.~Strominger, ``De Sitter space in
noncritical string theory,'' arXiv:hep-th/0205316;
E.~Silverstein, ``(A)dS backgrounds from asymmetric
orientifolds,'' arXiv:hep-th/0106209.
}

\lref\achar{
B.~S.~Acharya, ``A moduli fixing mechanism in M theory,''
arXiv:hep-th/0212294.
}

\lref\GKP{
S.~B.~Giddings, S.~Kachru and J.~Polchinski, ``Hierarchies from
fluxes in string compactifications,'' Phys.\ Rev.\ D {\bf 66},
106006 (2002) [arXiv:hep-th/0105097].
}

\lref\W{
E.~Witten, ``Anti-de Sitter space and holography,'' Adv.\ Theor.\
Math.\ Phys.\  {\bf 2}, 253 (1998) [arXiv:hep-th/9802150].
}

\lref\StromingerSH{ A.~Strominger and C.~Vafa, ``Microscopic
Origin of the Bekenstein-Hawking Entropy,'' Phys.\ Lett.\ B {\bf
379}, 99 (1996) [arXiv:hep-th/9601029].
}

\lref\GKPads{
S.~S.~Gubser, I.~R.~Klebanov and A.~M.~Polyakov, ``Gauge theory
correlators from non-critical string theory,'' Phys.\ Lett.\ B
{\bf 428}, 105 (1998) [arXiv:hep-th/9802109].
}

\lref\KachruNS{ S.~Kachru, X.~Liu, M.~B.~Schulz and S.~P.~Trivedi,
``Supersymmetry changing bubbles in string theory,'' JHEP {\bf
0305}, 014 (2003) [arXiv:hep-th/0205108].
}

\lref\GH{
G.~W.~Gibbons and S.~W.~Hawking, ``Cosmological Event Horizons,
Thermodynamics, And Particle Creation,'' Phys.\ Rev.\ D {\bf 15},
2738 (1977).
}

\lref\DS{
M.~Fabinger and E.~Silverstein, ``D-Sitter space: Causal
structure, thermodynamics, and entropy,'' arXiv:hep-th/0304220.
}

\lref\juan{
J.~M.~Maldacena, ``The large N limit of superconformal field
theories and supergravity,'' Adv.\ Theor.\ Math.\ Phys.\ {\bf 2},
231 (1998) [Int.\ J.\ Theor.\ Phys.\  {\bf 38}, 1113 (1999)]
[arXiv:hep-th/9711200].
}

\lref\dSCFT{
A.~Strominger, ``The dS/CFT correspondence,'' JHEP {\bf 0110}, 034
(2001) [arXiv:hep-th/0106113]
}

\lref\dSObj{
E.~Witten, ``Quantum gravity in de Sitter space,''
arXiv:hep-th/0106109;
W.~Fischler, A.~Kashani-Poor, R.~McNees and S.~Paban, ``The
acceleration of the universe, a challenge for string theory,''
JHEP {\bf 0107}, 003 (2001) [arXiv:hep-th/0104181];
S.~Hellerman, N.~Kaloper and L.~Susskind, ``String theory and
quintessence,'' JHEP {\bf 0106}, 003 (2001)
[arXiv:hep-th/0104180];
N.~Goheer, M.~Kleban and L.~Susskind, ``The trouble with de Sitter
space,'' JHEP {\bf 0307}, 056 (2003) [arXiv:hep-th/0212209].
L.~Dyson, J.~Lindesay and L.~Susskind, ``Is there really a de
Sitter/CFT duality,'' JHEP {\bf 0208}, 045 (2002)
[arXiv:hep-th/0202163];
T.~Banks, W.~Fischler and S.~Paban, ``Recurrent nightmares?:
Measurement theory in de Sitter space,'' JHEP {\bf 0212}, 062
(2002) [arXiv:hep-th/0210160];
T.~Banks and W.~Fischler, ``M-theory observables for cosmological
space-times,'' arXiv:hep-th/0102077.
}

\lref\stringstalk{E. Silverstein, talk at Strings 2003}

\lref\us{M. Fabinger, S. Hellerman, E. Silverstein, and others, in
progress.}

\lref\KLT{
P.~Kraus, F.~Larsen and S.~P.~Trivedi, ``The Coulomb branch of
gauge theory from rotating branes,'' JHEP {\bf 9903}, 003 (1999)
[arXiv:hep-th/9811120].
}

\lref\AbbottQF{ L.~F.~Abbott, ``A Mechanism For Reducing The Value
Of The Cosmological Constant,'' Phys.\ Lett.\ B {\bf 150}, 427
(1985).
}

\lref\BanksMB{ T.~Banks, M.~Dine and N.~Seiberg, ``Irrational
axions as a solution of the strong CP problem in an eternal
universe,'' Phys.\ Lett.\ B {\bf 273}, 105 (1991)
[arXiv:hep-th/9109040].
}

\lref\BrownKG{ J.~D.~Brown and C.~Teitelboim, ``Neutralization Of
The Cosmological Constant By Membrane Creation,'' Nucl.\ Phys.\ B
{\bf 297}, 787 (1988).
}

\lref\FengIF{ J.~L.~Feng, J.~March-Russell, S.~Sethi and
F.~Wilczek, ``Saltatory relaxation of the cosmological constant,''
Nucl.\ Phys.\ B {\bf 602}, 307 (2001) [arXiv:hep-th/0005276].
}


\Title{\vbox{\baselineskip12pt\hbox{hep-th/0308175}\hbox{SLAC-PUB-10142}
\hbox{SU-ITP-03/22}\hbox{NSF-KITP-03-70} }}{AdS and dS Entropy
from String Junctions}

\centerline{or}

\centerline{\bf The Function of Junction
Conjunctions\foot{Apologies to {\it Schoolhouse Rock} circa 1973,
http://www.schoolhouserock.tv/Conjunction.html}}

\bigskip

\centerline{Eva Silverstein\footnote{$^*$} {SLAC and Department of
Physics, Stanford University, Stanford, CA 94309}}

\vskip .3in \centerline{\bf Abstract} {Flux compactifications of
string theory exhibiting the possibility of discretely tuning the
cosmological constant to small values have been constructed. The
highly tuned vacua in this discretuum have curvature radii which
scale as large powers of the flux quantum numbers, exponential in
the number of cycles in the compactification. By the arguments of
Susskind/Witten (in the AdS case) and Gibbons/Hawking (in the dS
case), we expect correspondingly large entropies associated with
these vacua.  If they are to provide a dual description of these
vacua on their Coulomb branch, branes traded for the flux need to
account for this entropy at the appropriate energy scale. In this
note, we argue that simple string junctions and webs ending on the
branes can account for this large entropy, obtaining a rough
estimate for junction entropy that agrees with the existing rough
estimates for the spacing of the discretuum. In particular, the
brane entropy can account for the (A)dS entropy far away from
string scale correspondence limits.} \vskip .3in

\smallskip

\newsec{Introduction}

One of the most interesting recent developments is the
stabilization of moduli and construction of large classes of de
Sitter and anti de Sitter flux compactifications \MSS\achar\KKLT.
These models include cases in which the size of the
compactification is hierarchically smaller than that of the (A)dS,
by realizing the mechanism suggested in \BP\ (see also similar
works \refs{\AbbottQF,\BrownKG,\BanksMB,\FengIF}). The recent
models of KKLT \KKLT\ are of particular interest, as they produce
four dimensional de Sitter as well as anti de Sitter vacua in a
relatively well studied geometrical framework \GKP\ admitting a
low energy effective supersymmetric field theory
description.\foot{The space of models described in \MSS\ should be
taken into account in any attempt to bound the number of vacua,
and in comparing numbers of low energy SUSY vacua to vacua without
low energy SUSY. Ultimately it is quite possible that
nonsupersymmetric nongeometrical noncritical string backgrounds
may be more generic than the better studied geometrical low energy
SUSY backgrounds of critical string theory.}

It is of interest to look for holographic duals of these new flux
compactifications.  In the de Sitter case, such a description
could teach us a lot about the nature of dark energy (which in the
real world is roughly seventy percent of the observed universe) as
modeled in existing constructions.\foot{One may think about the
problem of dark energy in string theory analogously to the problem
of understanding black hole physics in string theory.  There is no
sense in which we try to ``explain" the black hole mass
independently of anything else, but we learn a lot about the
physics of black holes by understanding the microscopic origin of
their entropy \StromingerSH.}  Even in the AdS case the new
examples provide an interesting challenge. For four or fewer large
dimensions, previous nonperturbative formulations such as matrix
theory \bfss\ and AdS/CFT \juan\ examples obtained via near
horizon limits (such as $AdS_2\times S^2\times X$) have broken
down due to infrared problems.

Unlike the flux compactifications on large Einstein spaces which
have played a role in the AdS/CFT correspondence \juan, the new
examples are not (known to be) realized via a near horizon limit
of any simple brane systems. Nonetheless, there are general
arguments suggesting a similar holographic dual description.  In
the AdS case at least one expects a field theoretic dual via the
relation \GKPads\W\ mapping gravitational Feynman diagrams in
$AdS_{d+1}$ to conformally invariant Greens functions of a
$d$-dimensional quantum field theory.  This dictionary does not
depend on the existence of a larger theory from which the AdS
background is obtained as a near horizon limit.

In the dS case one also has a strong hint of a dual description,
in that the Gibbons-Hawking entropy \GH\ associated with the
horizon suggests a microphysical statistical mechanical origin
that may well be associated to a holographic dual theory. Steps
toward such a duality proposal using analogies to AdS/CFT have
been made in \dSCFT\ based on symmetries and the structure of
quantum field theory in the global de Sitter geometry, and in \DS\
based on entropy counts and geometry of brane configurations
realizing motion on the Coulomb branch of (A)dS flux
compactications of string theory.  In \dSObj\ some important
issues were raised that need to be addressed in any duality
proposal in the dS case.  In \tometal\ some proposals based on
novel quantum gravity constructions have been made and
investigated.  In my view, our best hope for finding a dual
formulation if one exists is to study the workings of explicit
models.

In \DS\us, a method for obtaining the dual theories for flux
compactifications has been proposed, as summarized in
\stringstalk\ (see also \achar). The idea is simply to deform the
system to the Coulomb branch, which introduces explicit brane
domain walls \domainwalls\ whose worldvolume content corresponds
to that of the dual field theory on its Coulomb branch. For the
AdS case, the solutions obtained by trading all the flux for
branes in the infrared region of the geometry have the property
that the solution caps off in the infrared, eliminating the AdS
horizon. In the well understood AdS/CFT examples, the brane
degrees of freedom at the scale of the VEV in this solution \KLT\
account for the full set of degrees of freedom of the known dual
field theory. In general flux compactifications, we would like to
understand if this is the case.

In \DS, we noted that the Bousso Polchinski tuning available for
flux compactifications suggests dual field theories with entropy
that is much greater than quadratic in the flux (and therefore
brane) quantum numbers. This makes more pressing the question of
whether the branes in a generic Coulomb branch configuration can
account for such a large entropy when the (A)dS space is much
larger than string scale in size. (When the (A)dS space is string
scale in size, there is a ``correspondence point" (cf \HorowitzNW)
at which the brane entropy scales like that of the (A)dS space if
the brane entropy is quadratic in the flux quantum numbers \DS.)

In this note, we show that the best current estimates for the
number of flux vacua in the KKLT system \BP\SKunp\DS\Douglas\
agrees with a simple estimate of the number of degrees of freedom
available on the branes realizing the Coulomb branch of the
system.  That is, from an estimate of the number of flux vacua,
one obtains an estimate of the smallest cosmological constant and
therefore the largest (A)dS radius scale available in the models.
Translating this to an entropy using the Bekenstein/Hawking,
Susskind/Witten, and Gibbons/Hawking arguments, one can compare
the result to an estimate of the number of degrees of freedom
available on brane domain walls in the Coulomb branch
configuration.  The latter count requires the inclusion of string
junctions and webs.  We find that the two estimates agree within
their theoretical error bars, though both estimates are most
reliably considered as lower bounds. In this way we relate the
statistics of flux vacua with the statistical mechanics of
individual flux vacua.

We further present a heuristic explanation of why this comparison
works (in our case and in the original case of AdS/CFT on the
Coulomb branch) based on the Susskind Witten analysis of entropy
in AdS vacua as a function of energy.

This result supports the idea that one can figure out the dual
theory from the information about its Coulomb branch available
directly on the gravity side, part of a program to determine the
duals under current development \us\ (see also
\refs{\achar,\AcharyaII}).\foot{Other aspects of the analysis \us\
include the relation between the vacua with fixed moduli on the
gravity side and the structure of renormalization group fixed
points on the field theory side, constraints on the quantum
numbers on the two sides,  and their structure under monodromies
of the compactification.} It improves our understanding of the
(A)DS entropy discussed in \DS\ for the cases in which the
cosmological constant is tuned to be very small.

This note is organized as follows.  In \S2\ we review the Bousso
Polchinski style estimate for the number of KKLT flux vacua.  In
\S3\ we review the deformation of the system to the Coulomb branch
via brane domain walls and present our estimate for the number of
degrees of freedom of the dual theory visible on the branes.  We
also present a heuristic explanation of the agreement between \S2\
and \S3\ based on the Susskind Witten analysis.

\newsec{Statistics of Flux Vacua}

The Bousso Polchinski mechanism predicts exponentially many vacua
as a function of multiple input flux quantum numbers, as follows
\BP\SKunp\DS\Douglas. A systematic approach to the problem of
counting flux vacua was recently developed in \Douglas. The basic
idea is the following. One expects a limit on the strength of flux
quantum numbers from back reaction on the geometry.  There are
$b_3$ RR flux quantum numbers $Q_i,i=1,\dots,b_3$ and $b_3$ NS
flux quantum numbers $N_i, i=1,\dots, b_3$.  If one expresses the
expected limitation in the form
\eqn\Rlim{R^2\equiv \sum_{i=1}^{b_3} \gamma_iQ_i^2+ \alpha_iN_i^2
< R^2_{max}}
for some order one coefficients $\alpha_i$ and $\gamma_i$, then
one obtains a total number of vacua which is of order
\eqn\totvac{N_{vac}\sim {R_{max}^{2b_3}\over b_3!}}
from the volume of the sphere in flux space containing the fluxes
consistent with \Rlim.  (This assumes that each choice of flux
leads to of order one vacua.)

In the KKLT models, this estimate may be given in terms of the
quadratic form
\eqn\Ldef{L\sim \int_{CY} H\wedge F}
as follows. Dimensional reduction on a space with flux produces
contributions to the four dimensional effective potential from the
flux kinetic terms for the NS flux $H_{NS}$ and the Ramond flux
$F_{RR}$
\eqn\fluxkin{\Lambda_{flux}\sim \int_{CY}{1\over l_4^2}{g_s^4\over
V^2} \sqrt{g}(|F_{RR}|^2+{1\over g_s^2}|H_{NS}|^2).}
where we are in 4d Einstein frame and $V$ is the compactification
volume in string units. This contribution takes the form
\eqn\BPquad{ \Lambda_{flux}\sim
\sum_{i=1}^{b_3}(c_iQ_i^2+a_iN_i^2) }
where $a_i$ and $c_i$ are functions of the moduli, which in turn
depend on the fluxes, and $Q_i$ and $N_i$ are the RR and NSNS flux
quantum numbers on the 3-cycles in the compactification. (In
asymmetric orbifold models such as \MSS\ the dependence of
$a_i,c_i$ on the moduli is eliminated for the geometrical moduli
by using asymmetic orbifolding to freeze them at the string
scale.)

If we pick the maximum flux scale $R_{max}$ such that the
moduli-dependent coefficients $a_i$ and $c_i$ do not take extreme
values in the solutions to the equations of motion, then one can
relate $L$ to a positive definite quadratic form for each point on
the moduli space solving the equations of motion.

That is, in the no scale models \GKP\ appearing in KKLT, the
Gauss' law relation between $L\sim \int H\wedge F$ and orientifold
3-plane and D3-brane charge
\eqn\gauss{{1\over{2(2\pi)^4(\alpha')^2}}\int H\wedge F = {1\over
4}(N_{O3}-N_{\overline{O3}})-N_{D3}+N_{\overline{D3}}}
translates via supersymmetry into a relation between the
orientifold +D3-brane tension and $L$.  In a zero energy vacuum of
the no-scale approximation \GKP\ to the effective potential, this
tension $\int H\wedge F$ cancels the positive terms \BPquad\ in
the potential.  So for every solution to the equations of motion
we wish to consider, a relation of the form
\eqn\boundI{ \sum a_i N_i^2+c_i Q_i^2\sim L \le R_{max}^2 }
holds, with $a_i$ and $c_i$ order one coefficients that depend on
the fluxes. So rewriting $R_{max}^2$ as $L_{max}$ we can rewrite
\totvac\ as
\eqn\Lform{ N_{vac}\sim {L_{max}^{b_3}\over {b_3!}} }
By integrating the number of vacua solving the equations of motion
over the flux choices and moduli space with a suppression factor
introduced for large fluxes to take into account \boundI,
\Douglas\ found an estimate
\eqn\Dest{ N_{vac}\sim {(2\pi L_{max})^K\over{12\pi^nn!K!}} f(K)}
where $K=b_3$ is the number of independent complex fluxes in the
compactification, and where $n=b_3/2-1$ is the complex dimension
of the complex structure moduli space of the Calabi-Yau threefold
associated to the F theory compactification.   $f(K)$ is an
integral of flux-independent quantities over a fundamental domain
of the moduli space.

If we take these vacua to be distributed roughly uniformly between
cosmological constants of $\pm {1\over l_4^2}$ (where $l_4$ is the
four-dimensional Planck length), this predicts a minimum
cosmological constant of magnitude
\eqn\minlam{\Lambda_{min}\sim {1\over {l_4^2N_{vac}}}}
corresponding to a maximum curvature radius $L_{(A)dS}$ of order
\eqn\curvscale{(L^{max}_{(A)dS})^2\sim l_4^2 N_{vac}}
among the elements of the discretuum of vacua predicted by the
estimate \Lform\Dest.  This curvature scale in turn corresponds to
an entropy of order
\eqn\entnow{S_{max}\sim {(L^{max}_{(A)dS})^2\over {l_4^2}}\sim
N_{vac}}
 as we will review
in the next section.  Taking the vacua to be uniformly distributed
is a nontrivial assumption, since the vacua could instead
accumulate around some particular values of the cosmological
constant.  We will see that this naive assumption fits with what
we find for the entropy, though a much more thorough analysis of
the distribution of vacua will ultimately be required.

This estimate, which may ultimately prove accurate as a count of
the number of vacua, appears at least to be a lower bound on this
number.  For example, we expect more solutions to the equations
fixing the complex structure and dilation moduli at the no scale
level than the $DW=0$ solutions so far counted \alexetal. In
addition, when we saturate Gauss' law with some number of
threebranes as well as fluxes, the number of vacua of the
threebrane theory comes into play and has not yet been estimated
accurately while at the same time fixing the moduli.\foot{We thank
S. Kachru for this caveat.} There are almost certainly other
classes of vacua such as \MSS\ to be included in a full count as
well, though the corresponding entropies for these may be studied
independently.

\newsec{Statistical Mechanics of Flux vacua}

Given an (A)dS vacuum of radius $L_{(A)dS}$, we can associate a
maximal entropy of order $L_{AdS}^{d-2}/l_d^{d-2}$ to a region of
the spacetime contained in a 2-sphere of radius $L_{(A)dS}$.  In
the dS case, this is simply the Gibbons Hawking entropy \GH.  In
the AdS case, this follows from applying the Bekenstein/Hawking
entropy bound to AdS space, as was studied for AdS/CFT by Susskind
and Witten \SW.

We will apply the Susskind Witten analysis to the Coulomb branch
configurations of our flux vacua in the AdS case.  Let us first
briefly review their analysis, generalizing trivially from the
$AdS_5\times S^5$ context in which they applied it. One begins
with an $AdS_{d}/CFT_{d-1}$ dual pair, for which the CFT has
$n_{CFT}$ degrees of freedom and therefore of order
$E^{(d-2)n_{CFT}}$ states in its spectrum as a function of energy
scale $E$.  Cut off this theory at a scale of order
$1/(L_{CFT}\delta)$ for some $\delta < 1$, where $L_{CFT}$ is the
size of the sphere on which the CFT lives.  The corresponding
operation on the gravity side is to place an infrared cutoff in
global AdS at a sphere of area $L_{AdS}^{d-2}/\delta^{d-2}$
surrounding the origin. Since precise coefficients are not
obtained by this analysis, for simplicity we may take $\delta$
somewhat smaller than but of order 1, so that the cutoff restricts
us to of order one mode per degree of freedom on the $S^{d-2}$ on
which the CFT lives. The area of this $S^{d-2}$ in Planck units,
$L_{AdS}^{d-2}/l_d^{d-2}$, bounds the entropy that can fit inside
the sphere on the gravity side. Susskind and Witten checked that
this entropy is indeed $N^2$ in the gravity dual to the ${\cal
N}=4$ $U(N)$ super Yang-Mills theory, using the relations
$L_{AdS}\sim L_{S^5}\sim (g_sN)^{1/4}l_s$.

Said differently, the cutoff requires each degree of freedom to be
excited with energy at most of order $1/L_{AdS}$.  The total
energy allowed below the cutoff is then $E_T=n_{CFT}/L_{AdS}$.
From the corresponding gravity side cutoff at a sphere of area
$L_{AdS}^{d-2}$, we can independently identify this total energy
$E_T$ as the mass $M_{BH}^{(L_{AdS})}$ of the largest black hole
fitting within the region bounded by this area.  In the
$AdS_5\times S^5$ case, these two formulas for the energy scale of
the cutoff agree, once we identify $n_{CFT}$ with $N^2$.  This
result is consistent with a naive extrapolation of the weak
coupling relation $n_{CFT}\sim N^2$ into the strong 'tHooft
coupling regime.

This analysis keeps track of the moding of states on the sphere as
well as the total entropy, and it illustrates a basic aspect of
how the entropy is distributed in the AdS/CFT
duality\foot{emphasized for example by S. Shenker}: from the
cutoff on the sphere, allowing only of order one mode on the
$S^{d-2}$ for of each of the $n_{CFT}=N^2$ degrees of freedom, one
obtains the entropy which is numerically equal to one degree of
freedom per Planck area but organized as $n_{CFT}=N^2$ degrees of
freedom per $L_{AdS}^3$.

The Susskind Witten analysis just reviewed was in the global AdS
solution. We can apply it in the Poincare patch, corresponding to
the CFT on Minkowski space $M_{d-1}$. We do this by enforcing the
Bekenstein bound corresponding to black brane solutions extending
in the $M_{d-1}$ directions.  This leads again to $N^2$ degrees of
freedom per $L_{AdS}^{d-2}$ area along the $d-2$ spatial
directions of $M_{d-1}$.

We would like to see how many of the $n_{CFT}$ degrees of freedom
of the system become manifest on its Coulomb branch. Let us first
review how the Coulomb branch arises from the gravity side point
of view. It is obtained by introducing brane domain walls
separated radially from the horizon.  This reduces the flux in the
bulk region on the side of the brane toward the horizon (let us
call this the ``IR side" since it corresponds to the IR region
from the field theory point of view). The simplest such
configuration, obtained in \KLT\ for the $AdS_5\times S^5$ case,
is to trade all the flux in this region for branes at a radial
scale of order $L_{AdS}$. There being no flux supporting the
compactification on the IR side of the branes, it shrinks down and
caps off the solution at a finite radius in the IR direction,
removing the horizon. In the solution \KLT, this region turns out
to be smooth (in fact flat) ten dimensional space.  In a general
flux compactification, we do not know the precise solution but the
absence of flux in this region means that the AdS horizon will be
removed generically. This corresponds to the fact that a generic
Coulomb branch configuration will lift most of the degrees of
freedom of the theory to a scale of order the scale set by the
VEVs.  Also in a generic system there will be a potential on the
Coulomb branch, so that the physical solutions are time dependent.
I expect this will not preclude the counting and identification of
degrees of freedom from the brane content on the gravity
side.\foot{Another work that used off shell bubbles to illustrate
a physics point is \KachruNS.}

In the $AdS_5\times S^5$ case, we see $N^2$ degrees of freedom
from stretched strings (``W bosons") at the mass scale
\eqn\vevscale{\langle\phi\rangle\sim {L_{AdS}\over l_s^2}}
of the VEVs of the diagonal scalar matrix elements.  (There are
also string oscillation modes on top of these including some at of
order this energy scale, which have the same $N$ scaling.) These
are electric degrees of freedom from the point of view of the
spontaneously broken $U(N)$ gauge group on the manifest D3-branes
of the \KLT\ solution, and become massless as we return to the
origin of the moduli space.  In this sense, the $N^2$ degrees of
freedom have become manifest on the Coulomb branch directly on the
gravity side of the correspondence.

Let us clarify the energy scales involved in this analysis. We put
the Susskind Witten cutoff originally at the radius $L_{AdS}$
corresponding to the total energy scale $N^2/L_{AdS}$. Exciting
the stretched string ``W bosons" individually fits within this
cutoff, so we can exhibit the count of degrees of freedom by
exciting them individually.  But exciting all $N^2$ of them at the
scale \vevscale\ would of course not fit inside the above cutoff,
which as we discussed allows states up to to a total energy scale
corresponding to $N^2$ degrees of freedom each excited only up to
energy $1/L_{AdS}$. So if one wants to apply the Susskind Witten
analysis to the system on the Coulomb branch, including energies
up to the scale
\eqn\Ecbr{ E_{C~branch}\sim N^2\langle\phi\rangle}
we need a larger cutoff (smaller $\delta$).



We can now ask in the more general flux compactifications of
interest here \KKLT\MSS\ whether the brane degrees of freedom
continue to account for the black hole entropy. That is, when we
count the elementary degrees of freedom $n_B$ on the brane domain
walls replacing all the flux to the IR end of the branes, at the
mass scale of the VEV suggested by the geometry, is $n_B$ of order
$L_{AdS}^{d-2}/l_d^{d-2}$? We will see that for the KKLT models,
at the level of the estimate in \S2\ this saturation holds as well
in our case, when we take into account string junction degrees of
freedom living on the branes in the Coulomb branch of that system.


The branes in the KKLT construction consist of $D5$ and $NS5$
branes wrapped on $b_3$ 3-cycles of the compactification manifold
of type IIB string theory, as well as of order $\int H\wedge F$
D3-branes ending on them according to the Gauss' law constraint
\gauss. $Q_i$ $D5$ and $N_i$ $NS5$ branes wrapped on the same
cycle $C_i$ reduce to $J_i$ $(p_i,q_i)$ fivebranes where
$(p_i,q_i)=(Q_i/J_i,N_i/J_i)$ are relatively prime integers. We
will be interested in the highly tuned situation described in \S2\
in which the size of the Calabi-Yau is much smaller than the
curvature radius of the $AdS_4$.  In particular, let us consider
all the length scales within the Calabi Yau to be somewhat bigger
than string scale for control but not parametrically bigger as a
function of the flux quantum numbers. Similarly, let us consider a
situation with $g_s$ somewhat smaller than one but of order one.

The degrees of freedom on the branes consist of strings and string
webs (combinations of string junctions) which are at a mass scale
of order
\eqn\vevscaleus{ m_{\langle\phi\rangle}\sim {1\over l_s} }
which is the analogue of \vevscale\ in our system. The string and
string web degrees of freedom can be electric from the point of
view of the gauge group on each bunch of branes. When the
classical mass and binding energy formulas are a good
approximation, some string webs are stable at an energy scale of
order \vevscaleus\ by virtue of being the lightest degrees of
freedom with their quantum numbers. We will estimate the number of
such degrees of freedom $n_B$ coming from string junctions that we
can reliably obtain ending on these various branes, and see that
they account for the entropy predicted on the gravity side \Dest -
\entnow.

%
\eqn\entpred{ n_B\sim N_{vac} }
with $N_{vac}$ given by \Dest.


We are interested in the number of degrees of freedom available on
the $N_5>>b_3$ 5-branes and $N_3>>b_3$ 3-branes obtained from the
$AdS_4$ solution by trading all the flux in the IR region of the
geometry for branes.
There will be multifundamental states arising from string webs
(connected combinations of string junctions) with multiple
external strings ending on the branes. String webs, discussed in
many interesting papers such as \strjunc\ (including one relating
them to black hole entropy \BL) are combinations of $(p,q)$
strings connected through three-string junction vertices. They
satisfy a basic charge conservation condition
\eqn\chargpq{ \sum_I (p_I, q_I)=0 }
where $I$ runs over the strings entering any vertex (and therefore
applies to the sum over external strings entering a string web).

We will start by studying junctions with one endpoint on each set
of branes (indexed by their type and by the cycle they wrap or end
on).  This will produce an entropy accounting for the gravity side
prediction.  We will not analyze the details of the flux
compactifications necessary to produce Coulomb branch
configurations with sufficiently stable solutions to \chargpq.  It
is clear that some such configurations exist, and our main goal is
to explain how large entropies of order $N_{vac}$ \entnow\Dest\
can arise in any Coulomb branch configuration of branes.

Before proceeding to the count of junctions, let us note two
issues we have not completely resolved. Firstly, junctions with
more endpoints on each bunch of branes, which could be viewed as
bound states of lower junctions, would lead to an entropy greater
than the gravity side prediction \Dest\entnow\ if such states were
considered separately. Secondly, the junctions we do consider with
one endpoint on each bunch of branes can themselves be viewed as
bound states of strings.  The question is what degrees of freedom
are elementary in the effective field theory at the energy scale
determined by the Coulomb branch VEVs.  These issues are similar
to those in a somewhat similar situation in the black hole context
in \BL, where just the lowest junction connecting three sets of
branes accounted for all the expected entropy. There is a
heuristic argument, along the lines of the arguments in \BL, that
higher bound state junctions are less likely to constitute valid
independent effective fields than the basic junctions with one
endpoint on each bunch of branes, since the ratio of the binding
energy of one constituent to the mass of the bound state decreases
as the number of endpoints increases. Because of this, the lowest
junctions are reliably counted but the higher ones are less and
less under control as we increase the number of endpoints, keeping
the size of the Calabi Yau fixed. Another argument due to \BL\ is
that in some cases the lowest junction states connecting different
bunches of branes ($U(N)$ factors in the brane system gauge group)
can be related to elementary string states in dual quiver
theories. These are not proofs that only the lowest junctions need
be considered however, and leaves open the possibility that {\it
more} entropy is available in the system than the naive Bousso
Polchinski tuning predicts. In any case, we will account for at
least the \BP\Douglas\ estimate with the simplest controlled
junction states, addressing the puzzle raised in \DS.

Let us study the junctions ending on the 3-branes and the
fivebranes. First consider the endpoints on the 3-branes. Consider
a generic situation where the ends of these 3-branes are
distributed roughly uniformly over the $b_3/2$ pairs ofdual
intersecting A and B cycles contributing to the anomalous $\int
H\wedge F$ 3-brane charge. Because of Gauss' Law \gauss, there are
of order $N_3\sim L$ D3-branes, and so of order $L/(b_3/2)$ per
pair of dual A and B cycles. A junction with one end on each of
the $b_3/2$ groups of $2L/b_3$ D3-branes has
\eqn\threen{n_3\sim \biggl({2L\over b_3}\biggr)^{b_3/2}}
ways to end on the threebranes.  In our estimates we will account
for the $L$-dependence and not reliably keep track of the
prefactor's dependence on $b_3$ (which is much smaller than $L$ in
the regime of validity of the analysis), though the strongest
factorial dependence on $b_3$ evident in \Dest\ will arise
naturally also in our estimate.

If the junction also ends on the $(p,q)$ fivebranes in all
possible ways (again with a single endpoint per bunch of branes),
then there is another factor in the entropy coming from the
fivebranes, which we now compute. Since we have of order $N_5/b_3$
$(p,q)$ 5-branes per 3-cycle, we have of order $n_5\sim
(N_5/b_3)^{b_3}$ ways the endpoints can end on fivebranes.

Let us relate this to the quantity $L$ with respect to which the
gravity side estimate \Dest\ is expressed.
Noting that
\eqn\NB{2N_5\sim \sum_{i=1}^{b_3}(|Q_i|+|N_i|)}
and recalling from \S2\ that
$L\sim \sum_{i=1}^{b_3}(c_iQ_i^2+a_iN_i^2)$
and using the fact that on average $Q_i\sim N_i\sim N_5/b_3$, we
obtain the relation
\eqn\LNB{ L^{{1\over 2}}\sim {N_5\over b_3}\sqrt{2b_3} }
This translates the fivebrane factor in the entropy to
\eqn\fiven{n_5\sim (L^{1/2}/\sqrt{b_3})^{b_3}.}

Putting the threebrane and fivebrane factors together, we obtain
\eqn\totest{ S_{junctions}\sim n_3n_5\sim \biggl({L\over
b_3}\biggr)^{b_3} }
The gravity side estimate \Dest\ does not determine the function
of $K=b_3$ multiplying the $L^K/K!$ factor, though \Douglas\
offered some arguments that it was subdominant in its $K$
dependence to the factorial in the denominator. At this level,
\totest\ from the lowest junctions with charge on all the sets of
branes agrees with the gravity side estimate \Lform\Dest\entnow.

Incidentally, one obtains the same estimate if one considers
junctions ending only on 5-branes if one decomposes the $(p,q)$
5-branes back into separate D and NS 5-branes.

Having recovered the entropy directly on the branes of our system,
a result similar to that obtained above for the ${\cal N}=4$ SYM
theory on its Coulomb branch, it is interesting to ask if this is
a coincidence or should have been expected.  The following is a
heuristic argument for the agreement based on the above Susskind
Witten analysis on the Coulomb branch.

Let us first consider the ${\cal N}=4$ super Yang-Mills theory. If
we consider the gravity side geometry out on the Coulomb branch in
a configuration in which all of the IR flux has been traded for
branes \KLT, this corresponds to a field theory configuration in
which the off diagonal matrix degrees of freedom have been lifted
to the scale $m_{\phi}$ of the VEVs.  The density of states of the
system for energies $E<N^2m_{\phi}$ scales like
\eqn\densIR{ \Omega(E)_{C branch}\sim E^{cN}}
which is much slower growth with energy than the undeformed CFT
scaling like
\eqn\densCFT{ \Omega_{CFT}(E)\sim E^{c^\prime N^2} }
for some constants $c, c^\prime$.  This means that in the dual
\KLT\ solution for the Coulomb branch, the black hole solutions
saturating the entropy at a given energy $E_0<N^2m_{\phi}$ contain
parametrically fewer states (i.e. have entropy of order $N$ rather
than $N^2$) than those in the full $AdS_5\times S^5$ geometry at
the same energy scale $E_0$.

If we now move the VEV scale $m_{\phi}$ down in energy to somewhat
below $E_0/N^2$, so that the branes go behind the black hole
horizon, then the two solutions (pure AdS and Coulomb branch)
agree for energies above $E_0$. In particular, the branes
contribute enough entropy to enhance the Coulomb branch black hole
density of states to \densCFT\ at energy $E_0$ rather than just
\densIR. This makes it clear why the brane states saturated the
order $N^2$ entropy (up to order one factors we do not control by
these considerations).

Now in the more general flux compactifications of interest here,
we are again applying the procedure of trading all the IR flux for
branes.  This again removes the flux stabilizing the
compactification, and probably caps off the solution in the IR.
This again suggests that the black holes in the capped off Coulomb
branch solution will have parametrically fewer states than in the
full AdS geometry at a given energy scale.  As in the above
discussion of the $AdS_5$ case, pushing the branes back behind the
horizon will produce again a black hole saturating the entropy
bound up to the energy scale $E_0$.  I therefore find it very
plausible that the branes in a KLT-like Coulomb branch
configuration in a general flux compactification will saturate the
entropy and will provide a reliable indicator of the content of
the holographic dual theory.  This bolsters considerably the case
for obtaining the content of the dual quantum field theory from
the branes on the Coulomb branch of the background \us.

There is a simple lesson from this analysis regarding the
distribution of the entropy. As emphasized above, in the Susskind
Witten analysis in ordinary AdS/CFT, the entropy is organized as
$n_{CFT}$ degrees of freedom per $L_{(A)dS}$ area. Both in AdS and
in dS flux compactifications, one can obtain numerological
agreement with the expected entropy in a situation where the
entropy is organized into one mode per {\it string} area per
intrinsic degree of freedom, rather than being organized into
$n_{CFT}$ degrees of freedom each excited by one mode per
$L_{(A)dS}^{d-2}$ area (as discussed in \S7\ of \DS). This
estimate is based on there being of order $Q^2$ degrees of freedom
in a system with $Q$ branes coming from open string degrees of
freedom. In this paper, we have seen that because their number
scales like larger than quadratic powers of the flux (brane)
quantum numbers, junction states can account for the expected
entropy $n_{junction}\sim N_{vac}$, arranged in the expected way
as $n_{junction}$ states per $(A)dS$ area rather than as one per
string area.

It will be very interesting to see if and how refinements of the
statistical analysis (keeping track of the specific configurations
required to tune the cosmological constant to be very small)
continue to lead to agreement between the two sides of the
putative duality.  In this note we have not addressed any aspect
of the distribution of flux vacua admitting large numbers of
degrees of freedom, but have only seen that it is possible and
very natural for string junction states to account for the large
entropy predicted for some vacua by the Bousso Polchinski
mechanism.

\smallskip
\bigskip
\centerline{\bf Acknowlegements}

I am very grateful to M. Fabinger and S. Hellerman, as well as X.
Liu, for useful discussions and collaboration on the larger
program \us\DS.  I would also like to thank S. Kachru for many
explanations of the workings of Calabi Yau flux compactifications.
In addition I have benefited from helpful comments on (A)dS
entropy from R. Bousso, M. Kleban, S. Shenker, and L. Susskind and
on string junctions from D. Berenstein, O. de Wolfe, and R. Leigh.
I would like to thank the hospitality of Strings 2003, the
Benasque Center for Science, the Globe Bookstore in Prague, the
Kavli Institute for Theoretical Physics, and Air Canada, KLM,
Czech Airlines, and Lufthansa for hospitality while this work was
carried out. Financial support comes from the DOE under contract
DE-AC03-76SF00515 and by the NSF under contract 9870115.  This
research was supported in part by the National Science Foundation
under Grant No. PHY99-07949.

\listrefs

\end